\documentclass[12pt]{article}
\usepackage[dvips]{color}
\usepackage{mathtext}
\usepackage[koi8-r]{inputenc}
\usepackage{graphicx}
\usepackage[figuresright]{rotating}
\usepackage{amssymb,eucal}
\usepackage{cite}
\date{}

\newcommand{\beq}{\begin{equation}}
\newcommand{\eeq}{\end{equation}}
\newcommand{\beqn}{\begin{eqnarray}}
\newcommand{\eeqn}{\end{eqnarray}}  

\begin{document} 
\title{Further search for T-violation in the decay $K^+\to \pi^0\mu^+\nu$ }

\author{M. Abe$^a$,
M. Aliev$^b$, V. Anisimovsky$^b$, M. Aoki$^c$, Y. Asano$^a$, \\
T. Baker$^d$, M. Blecher$^e$, P.~Depommier$^f$, M. Hasinoff$^g$, \\
K. Horie$^h$,  Y. Igarashi$^h$, J. Imazato$^h$,  A. Ivashkin$^b$, \\
M.~Khabibullin$^b$,  A. Khotjantsev$^b$, Yu. Kudenko$^b$,  Y. Kuno$^c$, \\ 
K.S. Lee$^i$, A. Levchenko$^b$,  G.Y. Lim$^h$,  J. Macdonald$^j$, \\
O. Mineev$^b$, N.~Okorokova$^b$, C.~Rangacharyulu $^d$, S.~Shimizu$^c$, \\ 
Y.-M.~Shin$^d$, N.~Yershov$^b$,  and T. Yokoi$^h$ \\
 ~\\ 
({\it Presented by Yu.~Kudenko for the KEK E246 Collaboration})
 ~\\
 ~\\
$^a$University of Tsukuba, 305-0006, Japan \\  
$^b$Institute for Nuclear Research RAS, 117312 Moscow, Russia \\  
$^c$Osaka University, 560-0043, Japan \\
$^d$University of Saskatchewan, S7N 0W0, Canada \\
$^e$Virginia Polytechnic Institute \& State University, USA \\
$^f$University of Montreal, H3C 3J7, Canada \\
$^g$University of British Columbia,  Vancouver V6T 1Z1,  Canada \\
$^h$KEK, 305-0801, Japan \\
$^i$Korea University, Seoul 136-701, Korea \\
$^j$TRIUMF, V6T 2A3,  Canada \\
}

\maketitle

\begin{abstract}
A new result for the transverse $\mu^+$ polarization, $P_T$,
in the decay  $K^+\to \pi^0\mu^+\nu$ has been obtained in the KEK E246 experiment. 
Combined with our earlier result, 
$P_T = (-1.12\pm 2.17(stat)\pm 0.90(syst))\times 10^{-3}$ and 
Im$(\xi)= (-0.28\pm 0.69(stat) \pm 0.30(syst))\times 10^{-2}$, which are 
consistent with no T-violation.
\end{abstract}

\section{Introduction}
The transverse muon polarization, $P_{T}$, 
in the decay $K^{+}\rightarrow\pi^{0}\mu^{+}\nu$~($K_{\mu3}$)
  provides a 
good opportunity to search for
 CP-violation beyond the Standard Model~(SM), and it can provide insight into the 
 origin of CP-violation. This polarization 
vanishes in the SM~\cite{bigi}, 
but it can be as large as $10^{-2}-10^{-3}$ in models with multi-Higgs doublets, 
leptoquarks, left--right symmetry or SUSY~\cite{geng}. Since  the 
contribution to  $P_{T}$ from  final 
 state interactions was found to be  $ < 10^{-5}$~\cite{fsi},  a larger  
 value of $P_T$ would be
a clear indication of physics beyond the SM  by inferring a non-zero 
Im$(\xi)$, where
$\xi(q^2) = f_{-}(q^2)/f_{+}(q^2)$ is  the ratio of two form factors, 
$f_{\pm}(q^2)$   in the $K_{\mu3}$ decay matrix element~\cite{cabibbo}. 
The previous result, 
$P_T = (-4.2\pm 4.9(stat)\pm 0.9(syst))\times 10^{-3}$ 
and 
Im$(\xi)= (-1.3\pm 1.6(stat) \pm 0.3(syst))\times 10^{-2}$ was obtained 
in~\cite{prl99} for the 1996-97 data set.   
In this paper we present a new  result 
of one of the  analyses of the data collected in 1998-2000 combined with our 
previous published result. 
\section{Experiment}
\begin{figure}[htbp]
\centering\includegraphics[width=150mm,angle=0]{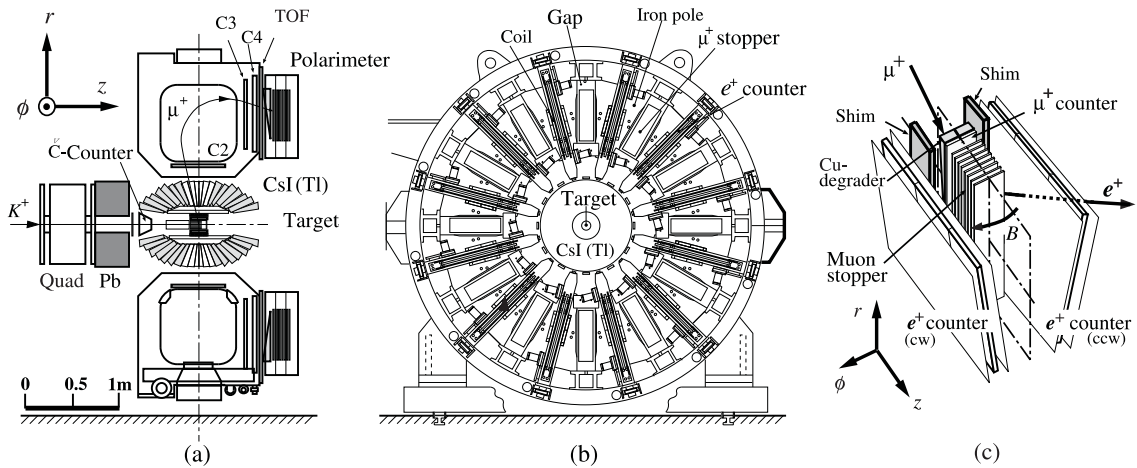}
\caption{Setup; (a) side view, (b) front view, and 
(c) one sector of the polarimeter.}
\label{fig:setup}
\end{figure}

The E246 experiment was carried out at the KEK 12-GeV proton synchrotron.
Detector  elements are described in Ref.~\cite{detector}.  
 In this experiment, the $K_{\mu3}$ decay of a stopped $K^+$ is identified 
 by detecting the
 $\pi^0$ as well as the $\mu^+$ from the decay. The  E246 
 setup is shown in Fig.~\ref{fig:setup}. 
 A 660-MeV/c  kaon beam is slowed down in a  degrader and stopped in a 
 scintillating fiber target.    
 The energy and direction of the $\pi^{0}$  from the $K_{\mu3}$ decay are 
 measured by a segmented CsI(Tl) photon detector installed in the central 
 region of a superconducting toroidal magnet. A muon from the 
 $K_{\mu3}$ decay at rest is momentum--analyzed in magnet gaps 
 by tracking in 
 the stopping target,  a scintillating ring 
 hodoscope surrounding 
 the target,  and three MWPCs (C2, C3, C4). The muon exiting the spectrometer
is stopped in a polarimeter
 in which the decay positron asymmetry $A_T$ is measured in order to obtain 
  $P_T$. The polarimeter consists of 12 azimuthally arranged Al stoppers, 
  aligned 
 with the magnet gaps, with scintillator counter system located between the 
 magnet gaps.
 A positron from the decay of $\mu^+\to e^{+}\nu\bar{\nu}$   is detected 
 by  plastic  scintillator counters located between the stoppers. 
 
In the new analysis, the $K_{\mu3}$  event selection 
is similar to~\cite{prl99}. The $\mu^+$ momentum region of 
100--190 MeV/$c$ was used to remove  the $K_{\pi2}$ decays. Most of the   
muons from pion decay in flight in  $K_{\pi2}$
are rejected by using the $\chi^2$ cut in tracking.
Neutral pions from the $K_{\mu3}$ decay are identified either  
 by $\gamma-\gamma$  coincidence in the    CsI and applying a   
 cut on the pion invariant mass~($2\gamma$ events),
 or by  
 one detected photon with large $E_{\gamma}$~($1\gamma$ events).
The $K_{e3}$  events which also satisfy these requirements  are 
removed by time-of-flight. In-flight kaon 
decays  were suppressed by  a   
requiring a delayed decay  after a $K^+$ is stopped.  
The ``good" $K_{\mu3}$ events were separated
into two classes:
$fwd$ events with the angle between $\pi^0 (\gamma)$ and beam direction  
$\theta_{\pi^0,\gamma} < 70^{\circ}$ and $bwd$ events with   
$\theta_{\pi^0,\gamma} > 110^{\circ}$.

The signal was extracted by integrating the positron time spectrum 
from
$\mu^+\to e^+\nu\bar{\nu}$ decays of muons stopped in the
polarimeter 
after subtraction of the background. 
The T-violating asymmetry $A_T$ is obtained as a difference in the counting
 rate between clockwise ($cw$) and counter-clockwise ($ccw$) emitted positrons. 
 Summing of the $cw$ ($N_{cw}$) and $ccw$ ($N_{ccw}$) positron counts  over 
 all 
 12 sectors, $A_T$ is derived from
\begin{equation}
A_T = \frac{1}{4} 
 \Bigl [ \frac{(N_{cw}/N_{ccw})_{fwd}}{(N_{cw}/N_{ccw})_{bwd}} - 1 \Bigr ].
\end{equation}
Then,  $P_T = A_T/(\alpha\cdot f)$,   
where the analyzing power of the polarimeter is $\alpha = 0.281\pm 0.015$, 
obtained from asymmetry  measurement
of the in-plane component of $\mu^+$ polarization, $P_N$, by selecting 
$\pi^0$s emitted transverse to the beam and comparing to a Monte Carlo
calculation.   
The kinematic attenuation factor $f$ results from accepting 
$fwd$ and $bwd$ events with $|{\rm cos}\theta_{\pi^0,\gamma}| > 0.342$ and 
was also 
obtained  from a Monte Carlo calculation. It has  different values for 
$1\gamma$ and $2\gamma$ events: $f = 0.72-0.77$ for $2\gamma$
and $f = 0.56-0.66$ for $1\gamma$ events, depending on the background 
level in the CsI. Then, 
${\rm Im}(\xi) = P_T/\Phi$, where $\Phi\simeq 0.33(0.29)$ for 
$2\gamma$ ($1\gamma$) events is a kinematic factor 
obtained from the analysis of the $K_{\mu3}$ Dalitz distribution.

The contamination of the beam accidental backgrounds in ``good"  $K_{\mu3}$ 
events was about 8\%~($2\gamma$),
$\sim 9$\%~($1\gamma$), and the constant background in the polarimeter 
was $11-12$\%. These backgrounds only diluted the sensitivity to $P_T$
by 10\%, but they did not produce any spurious T-violating asymmetry.
The main systematics uncertainties in $P_T$ come from the  two
large in-plane components  
of the $\mu^+$  polarization, $P_L$ which is 
parallel to
the muon momentum and $P_N$ ($P_T\ll P_{N,L}\leq 1$). 
The largest  systematic errors are due to  the  misalignment of
the polarimeter, the asymmetry of magnetic field distribution, and the
asymmetrical kaon stopping distribution. Most of these effects are canceled 
by the azimuthal symmetry of the detector 
as well as by  
the $fwd/bwd$ ratio. For example, the effect of the kaon stopping 
distribution is reduced by more than a factor of 10~\cite{kudenko02}. The total
systematic error of $P_T$ is estimated to be the same as that of 
the previous result~\cite{prl99}.

The data analysis was performed by two independent groups. Both analyses
obtained consistent results for 1996-97 data set (see Ref.~\cite{prl99}).   
The result 
for 1998-2000 data set was obtained by one of the  analyses.  
\section{Result} 
\begin{figure}[htb]
\centering\includegraphics[width=104mm,angle=0]{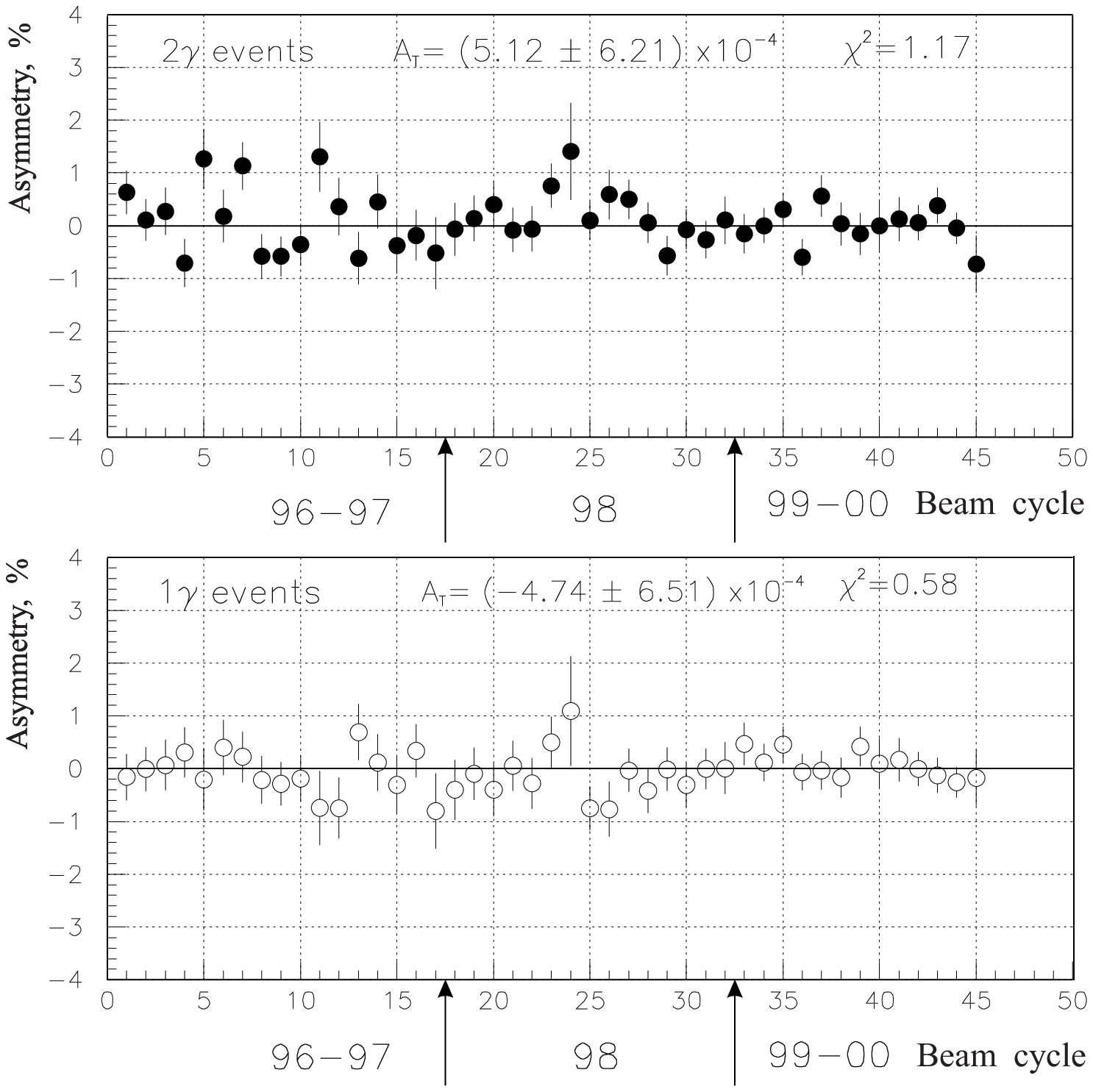}
\caption{T-violating asymmetry consistency check for $2\gamma$ and $1\gamma$ 
$K_{\mu3}$ events. 
Independent statistical errors are shown. Vertical arrows separate 
three long beam cycles. For 1996-97 cycle the results of one of the analyses are 
plotted.
 }
\label{fig:asymmetry}
\end{figure}

In 1998--2000 we selected
about  $4.4 \times 10^6$ ``good" 
$K_{\mu3}$  events. To check the stability of the result,  all 
data--taking periods, including 1996-97, were divided roughly in  
100-hour time intervals, and  $A_T$   
for $1\gamma$ and $2\gamma$ events in each interval were calculated. The 
results are presented 
in Fig.~\ref{fig:asymmetry}. 
The average asymmetries for both classes of events are consistent with zero
within a $\pm 1\sigma$ interval. 

Combining the new and previous 
results\footnote[1]{For the previous result the newly obtained value of $\alpha$ 
was also applied}, we obtain preliminary   
$P_T = (-1.12\pm 2.17(stat)\pm 0.90(syst))\times 10^{-3}$ 
and
Im$(\xi)= (-0.28\pm 0.69(stat) \pm 0.30(syst))\times 10^{-2}$,
 consistent with no T-violation in $K_{\mu3}$.

\end{document}